\documentclass[
 reprint,
 amsmath,amssymb,
 aps,
]{revtex4-2}

\usepackage{graphicx}
\usepackage{bm}

\usepackage[colorlinks=true, linkcolor=blue,anchorcolor=blue, citecolor=blue,urlcolor=blue]{hyperref}

\usepackage[normalem]{ulem}
\begin{document}
\preprint{APS/123-QED}

\title{Multistate Manipulation of Charge–Spin Conversion in Two-Dimensional Ferroelectric Bilayers}
\author{Weiyi Pan$^{1}$}
\email{Weiyi.Pan@physik.uni-regensburg.de}
\author{Xinyuan Jiang$^{2}$}
\author{Jaroslav Fabian$^{1,3}$}

\affiliation{$^{1}$Institute for Theoretical Physics, University of Regensburg, 93040 Regensburg, Germany\\
$^{2}$State Key Laboratory of Low Dimensional Quantum Physics, Department of Physics, Tsinghua University, Beijing 100084, China\\
$^{3}$Halle-Berlin-Regensburg Cluster of Excellence CCE, University of Regensburg, 93040 Regensburg, Germany\\
}

\begin{abstract}
Achieving nonvolatile and multistate manipulation of charge–spin conversion, including the Edelstein effect (EE) and spin Hall effect (SHE), is crucial for high-density spintronic memory. Here, we propose a mechanism to simultaneously control both EE and SHE in two-dimensional ferroelectric bilayers, where interlayer-parallel and interlayer-antiparallel polarization configurations can coexist. Symmetry analysis shows that in the interlayer-parallel states, reversal of the total polarization switches the sign of the EE, whereas changing to an interlayer-antiparallel configuration suppresses the EE to zero, enabling electrically switchable current-induced spin accumulation among three distinct states,  that could be used for ternary logic operations. Meanwhile, the magnitude of the SHE can be tuned by switching between two different classes of polarization configurations (interlayer-parallel and interlayer-antiparallel). Using first-principles calculations, we demonstrate this mechanism in bilayer metallic ferroelectric PtBi$_{2}$, where both interlayer-parallel and interlayer-antiparallel polarization configurations are energetically stable. The EE coefficient $\chi_{yx}$ in interlayer-parallel states, which reaches ${10}^9 - {10}^{10} \hbar/(\textup{A}\cdot \textup{cm})$ near the Fermi energy and can be reversed by polarization switching, arises from competing electron- and hole-pocket contributions near the Fermi surface. The intrinsic SHE coefficients, which are in the range of 100 - 1000 $(\hbar/e) \cdot (\textup{S/cm})$ near the Fermi energy, originate from spin Berry curvature that can be reshaped by polarization configuration variation and Fermi-level tuning. Our results establish ferroelectric bilayers as an all-in-one platform for electrically programmable charge–spin conversion.

\end{abstract}
\maketitle
\section{Introduction}
Charge–spin conversion is a fundamental phenomenon in spintronics\cite{RevModPhys.76.323}, enabling the mutual coupling between charge currents and spin degrees of freedom and thus playing an essential role in achieving all-electric control with low energy consumption\cite{puebla2020spintronic}. It is typically manifested in two representative effects: the Edelstein effect (EE)\cite{EDE,Johansson_2024}, which describes current-induced nonequilibrium spin accumulation in noncentrosymmetric systems\cite{REE1,REE2,REE3,REE4,REE5,REE6,REE7,REE8,REE9,REE10,REE11,REE12,REE13,REE14,REE15,REE16,REE17}, and the spin Hall effect (SHE)\cite{SHE}, which denotes the transverse spin current generated by a longitudinal charge current. Quantitatively, the EE can be described by
$\delta S_i=\chi_{ij}E_j$\cite{Xu,Johansson_2024},
where an applied electric field $E_j$ induces a nonequilibrium spin accumulation $\delta S_i$ along the $i$-direction. Meanwhile, the SHE can be described by
$J_j^i=\sigma_{jk}^iE_k$\cite{SHE,PhysRevLett.92.126603},
where an applied electric field $E_k$ generates a spin current flowing along the $j$-direction with spin polarization along the $i$-direction, with $i$, $j$ and $k$ being mutually orthogonal.

An important frontier in this field is the realization of nonvolatile electrical control of charge–spin conversion, which is essential for reconfigurable spintronics applications, including spin logic\cite{schenk2020memory,ESOT,ESOT2,ESOT3} and neuromorphic computing\cite{grollier2020neuromorphic}. Ferroelectric materials provide a promising platform for this purpose, because their electrically switchable spontaneous polarization can couple to spin–orbit-coupled electronic structures and thereby modulate charge–spin conversion\cite{picozzi2014ferroelectric,gu2024ferroelectric}. Consequently, charge–spin conversion in ferroelectric materials has attracted growing attention as a route to electrically switchable spin–orbit phenomena.

In many ferroelectric systems, polarization reversal is generally equivalent to a spatial inversion operation, resulting in a reversal of the sign of the EE. Such ferroelectrically switchable EE has been demonstrated in several systems based on ferroelectric semiconductors \cite{FE1,FE2,FE3,FE4,FE5,FE6,FE7}, including GeTe\cite{FE1,FE7}, In$_{2}$Se$_{3}$\cite{FE3}, graphene/In$_{2}$Se$_{3}$ heterostructure\cite{MIS}, and CsGeX$_{3}$ (X = I, Br, Cl)\cite{FE6}. 

A significant milestone was achieved with the discovery that 2D metals, despite electrostatic screening, can exhibit not only polarizations (predicted for bulk crystals \cite{FE9}  and observed in LiOsO$_3$ \cite{FE10}), but also ferroelectric switching, which was recently proposed \cite{Wu,PhysRevB.108.104109} and
experimentally demonstrated in the 1T' phase of few-layer WTe$_2$ \cite{FE9}. Ferroelectric control of EE, without external doping, has been studied
theoretically in bilayer 1T'-WTe$_{2}$\cite{FE4} and monolayer PtBi$_{2}$\cite{j5s5-m7j5}, predicting efficient switchable EE. 
In these studies, ferroelectric switching reverses the sign of the EE, thereby enabling binary-state control. 

If a ferroelectric system could realize nonvolatile control among multiple polarization states by an electric field, it would provide an opportunity for multistate modulation of the EE. For example, such a system could simultaneously achieve sign reversal and magnitude tuning of the EE, or enable switching among opposite EE signs and an on/off state, thereby allowing higher information storage density. Nevertheless, multistate manipulation of the EE in ferroelectric systems remains elusive.

A related challenge concerns the nonvolatile control of the SHE. But in contrast to the EE, the SHE conductivity is invariant under spatial inversion. As a result, polarization reversal, which is usually equivalent to a spatial inversion operation, cannot by itself modulate the SHE. Thus, various efforts have been devoted to manipulating the SHE in ferroelectric systems with alternative mechanisms. For instance, the SHE can be strongly modified by changing the atomic configuration from a ferroelectric phase to a paraelectric phase, which may be achieved by increasing the temperature\cite{wang2020spin,GeTe2}. However, such a transition cannot be realized nonvolatilely by an electric field, because the paraelectric phase is generally not stable under an external electric field. In addition, several studies have suggested nonvolatile electrical control of the SHE in heterostructures consisting of a ferroelectric substrate and a nonmagnetic metal\cite{fang2020tuning,xu2022regulating,chen2025quantum}. In these systems, the two states with opposite polarizations are generally not related by any symmetry operation, thereby allowing distinct SHE responses between opposite polarization states. Nevertheless, lattice mismatch at the interface of such heterostructures inevitably introduces defects, which may lead to uncontrolled deviations in the SHE behavior. Therefore, it is highly desirable to identify new physical mechanisms in intrinsic polar systems that enable nonvolatile electrical control of the SHE. 

In this work, we propose a new approach that enables simultaneous nonvolatile and multistate manipulation of both the EE and SHE in ferroelectric systems. We consider a bilayer two-dimensional (2D) ferroelectric system, in which the polarization state can be switched not only between interlayer-parallel configurations with opposite total polarizations, but also between interlayer-parallel and interlayer-antiparallel configurations. We show that, in such a ferroelectric bilayer system, switching between two distinct interlayer-parallel states with opposite total polarizations reverses the sign of the EE while leaving the SHE unchanged. In contrast, switching between an interlayer-parallel state and an interlayer-antiparallel polarization state not only enables on–off switching of the EE, but also realizes nonvolatile modulation of the SHE. We demonstrate this mechanism in bilayer PtBi$_{2}$, where switching among distinct polar states leads to modulation of both the EE and SHE. For EE in polarization-parallel states, its value reaches ${10}^9 \sim {10}^{10} \hbar/(\textup{A}\cdot \textup{cm})$, and shifting the Fermi level upwards would cause the sign reverse of EE. For SHE, it reaches the value of 100 - 1000 $(\hbar /e) \cdot$ (S/cm) and its magnitude can be modified by tuning the Fermi level due to the reshape of spin Berry curvature. Our findings provide further insight into the nonvolatile manipulation of charge–spin conversion and may be useful for future spintronic devices.

\section{Results and discussions}
\subsection{Design principle}
\begin{figure}[ht]
\includegraphics[scale = 0.22 ]{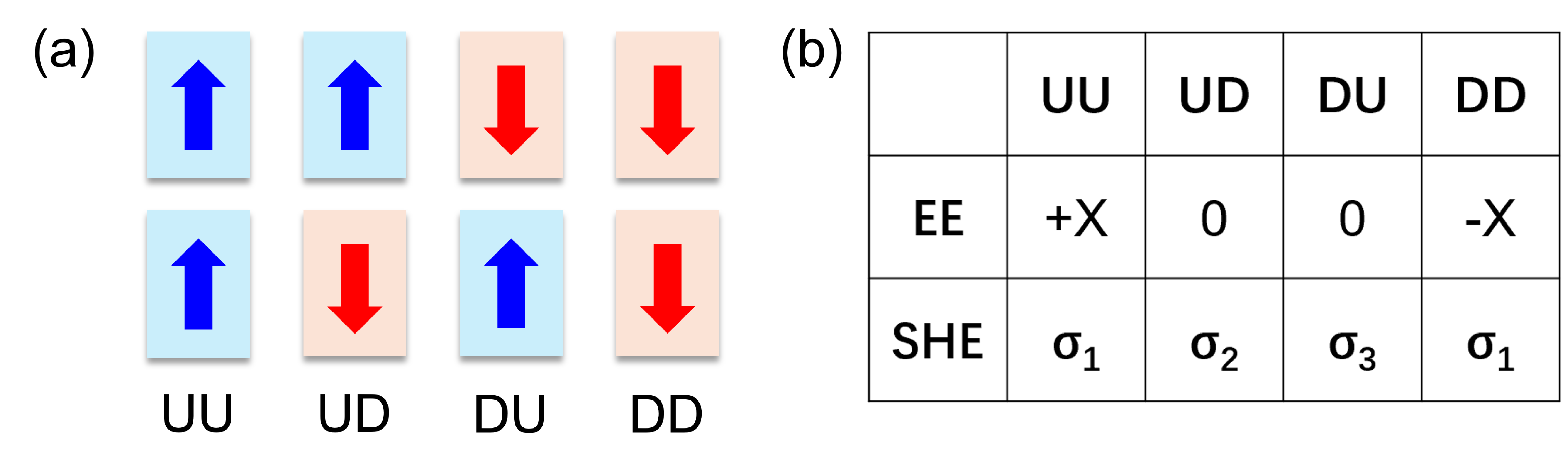}
\caption{\label{1} The design principle of multistate-tunable charge-spin conversion. (a) Schematic illustration of four polarization configurations in a bilayer ferroelectric systems. (b) The behaviour of EE and SHE under distinct polarization configurations. Here $X$ denotes the magnitude of EE, while $\sigma_{1}$, $\sigma_{2}$ and $\sigma_{3}$ denote three different magnitudes of SHE. }
\end{figure}

In a single-layer 2D ferroelectric system with out-of-plane polarization, an external electric field can switch the polarization between the up and down states, which are related by spatial inversion operation. Upon such polarization switching, the EE reverses its sign while retaining the same magnitude, whereas the SHE remains unchanged. The situation becomes more versatile when two layers of a 2D ferroelectric material are stacked to form a homobilayer, as illustrated in Fig. \ref{1}(a). In such a ferroelectric bilayer, both parallel and antiparallel alignments of the electric dipoles in the two layers are, in principle, allowed, giving rise to four polarization states: interlayer-parallel up–up (UU), interlayer-antiparallel up–down (UD), interlayer-antiparallel down–up (DU), and interlayer-parallel down–down (DD). Among these states, the interlayer-parallel ferroelectric-like configurations UU and DD possess finite polarizations with equal magnitude and opposite sign. Since they are related by spatial inversion operation, they host EEs with equal magnitude but opposite sign, while their SHEs are identical, as shown in Fig. \ref{1}(b). In contrast, the interlayer-antiparallel antiferroelectric-like UD and DU states are inversion-symmetric and therefore forbid a net EE. Meanwhile, because no symmetry operation relates the UU, UD, and DU states, their SHEs are not constrained by symmetry to be identical and can therefore differ from one state to another, as shown in Fig. \ref{1}(b). Based on the above analysis, one can infer that: (i) switching the polarization between the two interlayer-parallel UU and DD states enables sign reversal of the EE; and (ii) switching between the two classes of interlayer configurations: interlayer-parallel states (UU and DD) and the interlayer-antiparallel states (UD and DU), provides an effective route to realize both on–off control of the EE and nonvolatile modulation of the SHE. These results highlight the potential of 2D ferroelectric bilayers as an all-in-one platform for the nonvolatile manipulation of charge–spin conversion. 

\subsection{Basic structural and electronic properties}

Now we demonstrate the above design principle using 2D trigonal PtBi$_{2}$ as a representative example. Bulk trigonal PtBi$_{2}$ is accessible experimentally\cite{PhysRevB.94.165119,PhysRevB.94.235140,PhysRevMaterials.4.124202,changdar2025topological}, and a previous theoretical study has predicted that monolayer PtBi$_{2}$ can be exfoliated from bulk trigonal PtBi$_{2}$ and can exhibit an intrinsic metallic ferroelectric state with electrically reversible out-of-plane polarization\cite{Wu}. Moreover, the $C_{3v}$ point-group symmetry of monolayer PtBi$_{2}$ allows the EE component $\chi_{yx}=-\chi_{xy}$, whose sign can be reversed by ferroelectric switching\cite{j5s5-m7j5}. By considering different alignments of the electric polarization in each layer, four bilayer configurations can be constructed: two interlayer-parallel states with opposite net polarizations, namely UU and DD, which possess non-centrosymmetric $C_{3v}$ symmetry and therefore allow the existence of the EE; and two interlayer-antiparallel states without net polarization, namely UD and DU, which possess centrosymmetric $C_{2h}$ symmetry and therefore forbid the EE. By evaluating the transition pathways among these states [see Fig. \ref{7} in Appendix B], we find that the interlayer-parallel UU and DD states are energetically favored, whereas the interlayer-antiparallel UD and DU states are metastable configurations that can be accessed from the UU or DD states under an external electric field. Quantitatively, the calculated energy barriers for transforming from the UU/DD state to the UD and DU states are approximately 0.36 and 0.44 eV for each unit cell, respectively, while the barriers for transforming from the UD and DU states back to the UU or DD state are 0.23 and 0.12 eV for each unit cell, respectively. These energy barriers are comparable to those of several experimentally demonstrated 2D ferroelectric systems, such as CuCrSe$_{2}$ (0.19 eV)\cite{10.1093/nsr/nwz169}, bilayer MoS$_{2}$ (0.23 eV)\cite{PhysRevLett.112.157601}, and CuInP$_{2}$S$_{6}$ (0.2 eV)\cite{doi:10.1021/acs.jpclett.2c00105}.

 \begin{figure}[ht]
\includegraphics[scale = 0.21 ]{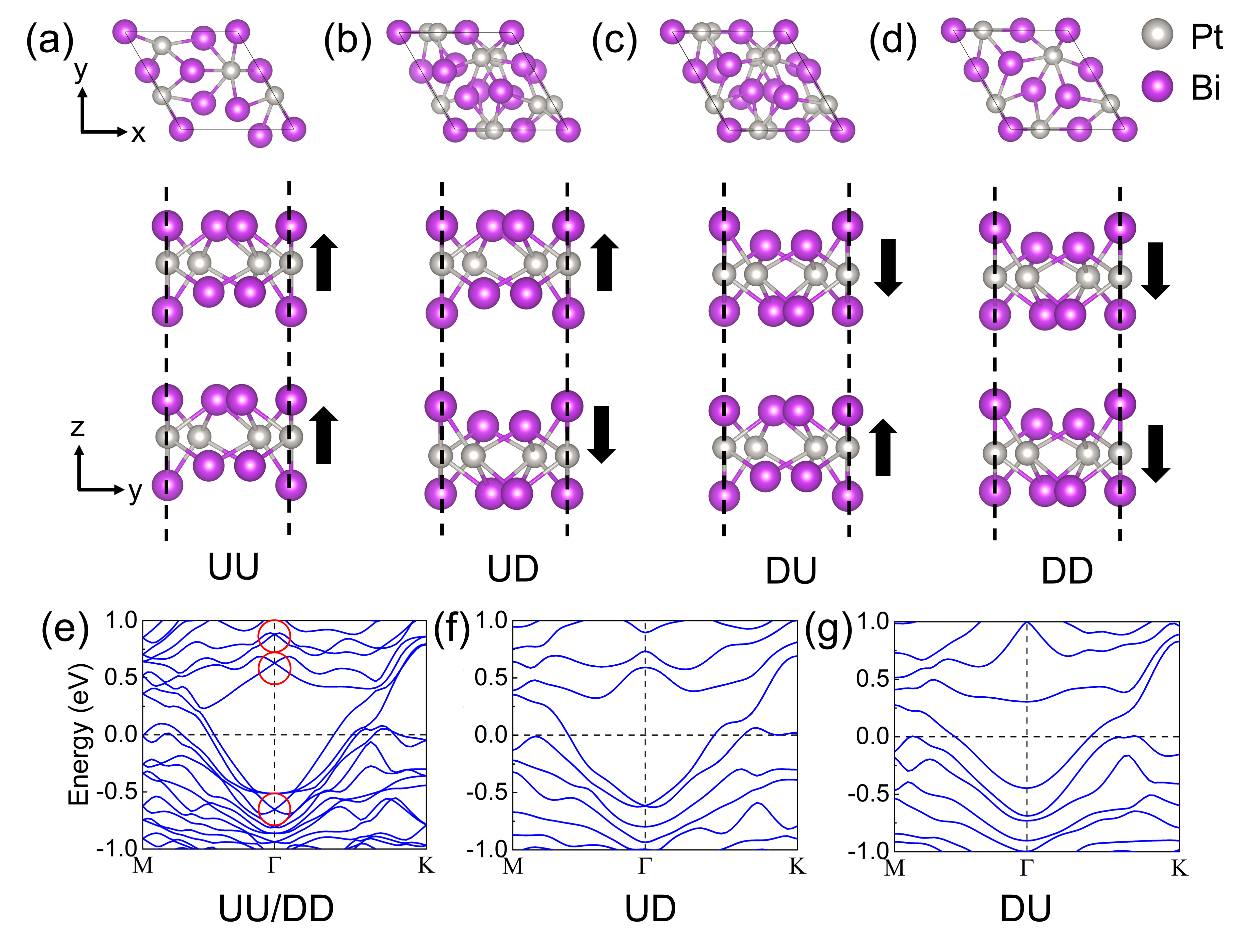}
\caption{\label{2} Structure and electronic properties of bilayer PtBi$_{2}$. (a)-(d) denotes the bilayer PtBi$_{2}$ with four distinct polarization configurations, namely, interlayer-parallel up–up (UU), interlayer-antiparallel up–down (UD), interlayer-antiparallel down–up (DU), and interlayer-parallel down–down (DD). The black arrow denotes the direction of electric polarization in each monolayer. (e)-(g) denotes the band structure of bilayer PtBi$_{2}$ in UU, UD and DU configuration, respectively.  In (e) the Rashba-like band splittings at $\Gamma$ point are indicated in the circles.}
\end{figure}

By changing the polarization state, the electronic band structure also changes, which is shown in Figs.\ref{2}(e)-\ref{2}(g). As can be seen, all the interlayer-parallel (UU/DD) and interlayer-antiparallel (UD/DU) states are metallic state. Note that both stable interlayer-parallel and interlayer-antiparallel states have so far been observed predominantly in 2D insulating ferroelectric systems\cite{PhysRevLett.126.057601,PhysRevLett.134.166401,fj5v-d4mz,PhysRevB.111.L081407,PhysRevLett.133.146605}, whereas their metallic counterpart remains exceptionally rare in experimentally realized materials, despite their great potential for high-efficiency, nonvolatile, and electrically controllable electronic devices. Specifically, for UU/DD state which breaks inversion symmetry, the band structure near the Fermi level is quite complicate, with various Rashba-like band splitting occurs at the $\Gamma$ point. Near the $\Gamma$ point, two bands exhibiting quadratic dispersion cross the Fermi level and form two electron pockets. In addition, several other bands intersect the Fermi level at momenta away from $\Gamma$ point and thus forms hole pockets. In contrast, the UD and DU states preserve inversion symmetry. Consequently, all bands remain spin degenerate under $PT$ symmetry, and Rashba-like spin splitting is forbidden. This absence of spin splitting gives rise to a less congested electronic band structure near the Fermi level, as shown in Fig. \ref{2}(f) and \ref{2}(g).

\subsection{Edelstein effect}
 \begin{figure}[ht]
\includegraphics[scale = 0.42 ]{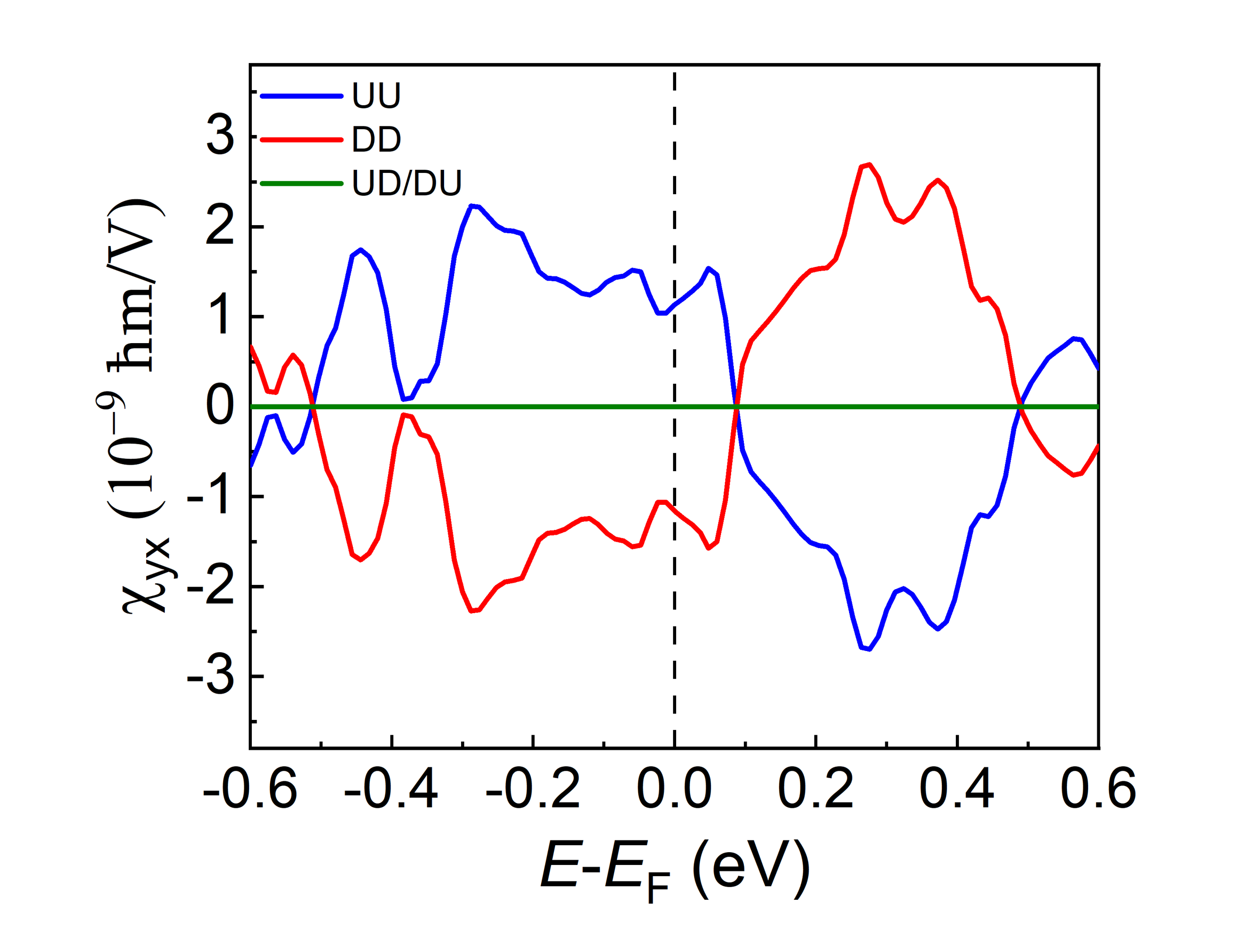}
\caption{\label{3} Calculated EE coefficient in bilayer PtBi$_{2}$ as a function of Fermi energy. Note that the EE in both UD and DU states are zero, which is guaranteed by the inversion symmetry. A broadening of $\Gamma$ = 0.01 eV is adopted. }
\end{figure}

 \begin{figure*}[ht]
\includegraphics[scale = 0.47 ]{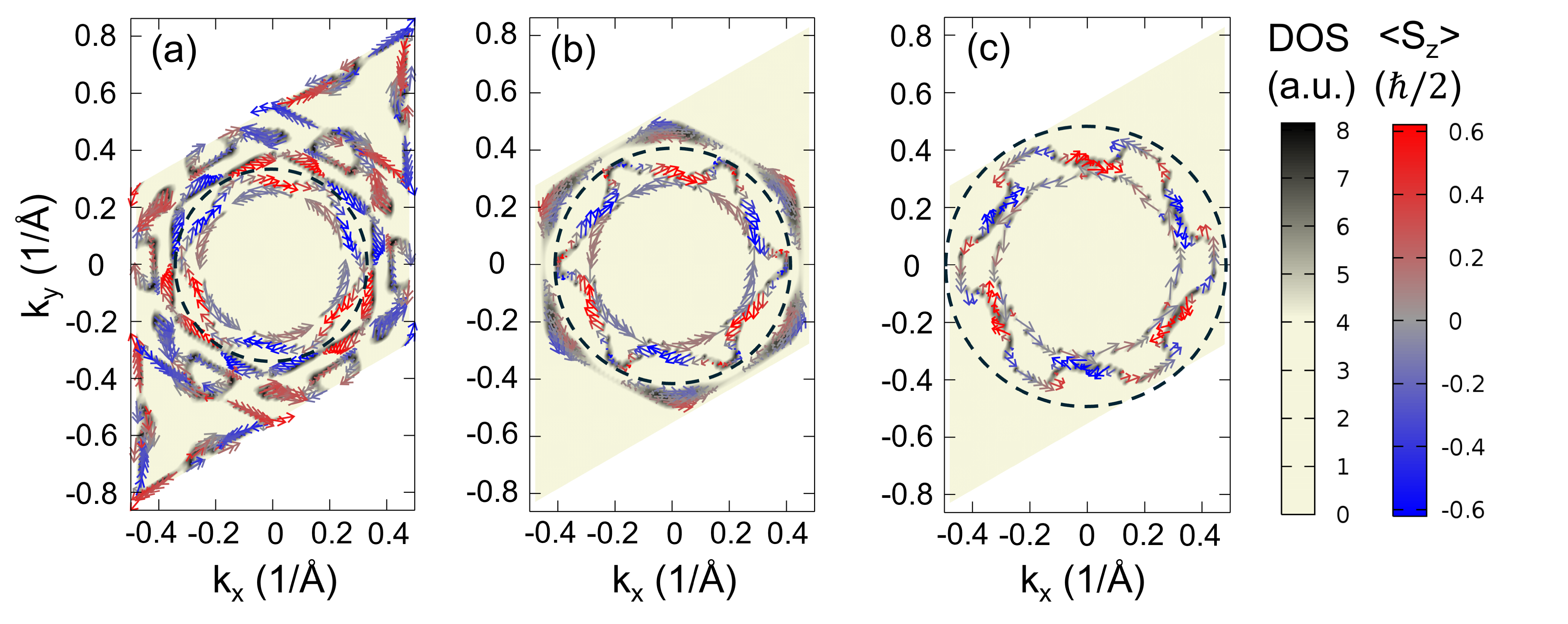}
\caption{\label{4} The Fermi surface states as well as the corresponding spin textures for UU state of bilayer PtBi$_{2}$ at (a) $E_{F}$ = 0 eV, (b) $E_{F}$ = 0.1 eV, and (c) $E_{F}$ = 0.2 eV, respectively. The color of background shows the spectral density of the electronic states on the Fermi surface, which is denoted in arbitrary unit. The arrows shows the direction and intensity of in-plane spin expectation values of these Fermi surface states. The color of arrow represents the expectation value of out-of-plane spin components. The states from inner electron pockets near the $\Gamma$ point are circled, while the states out of the circle are from outer hole pockets. }
\end{figure*}

 \begin{figure*}[ht]
\includegraphics[scale = 0.43 ]{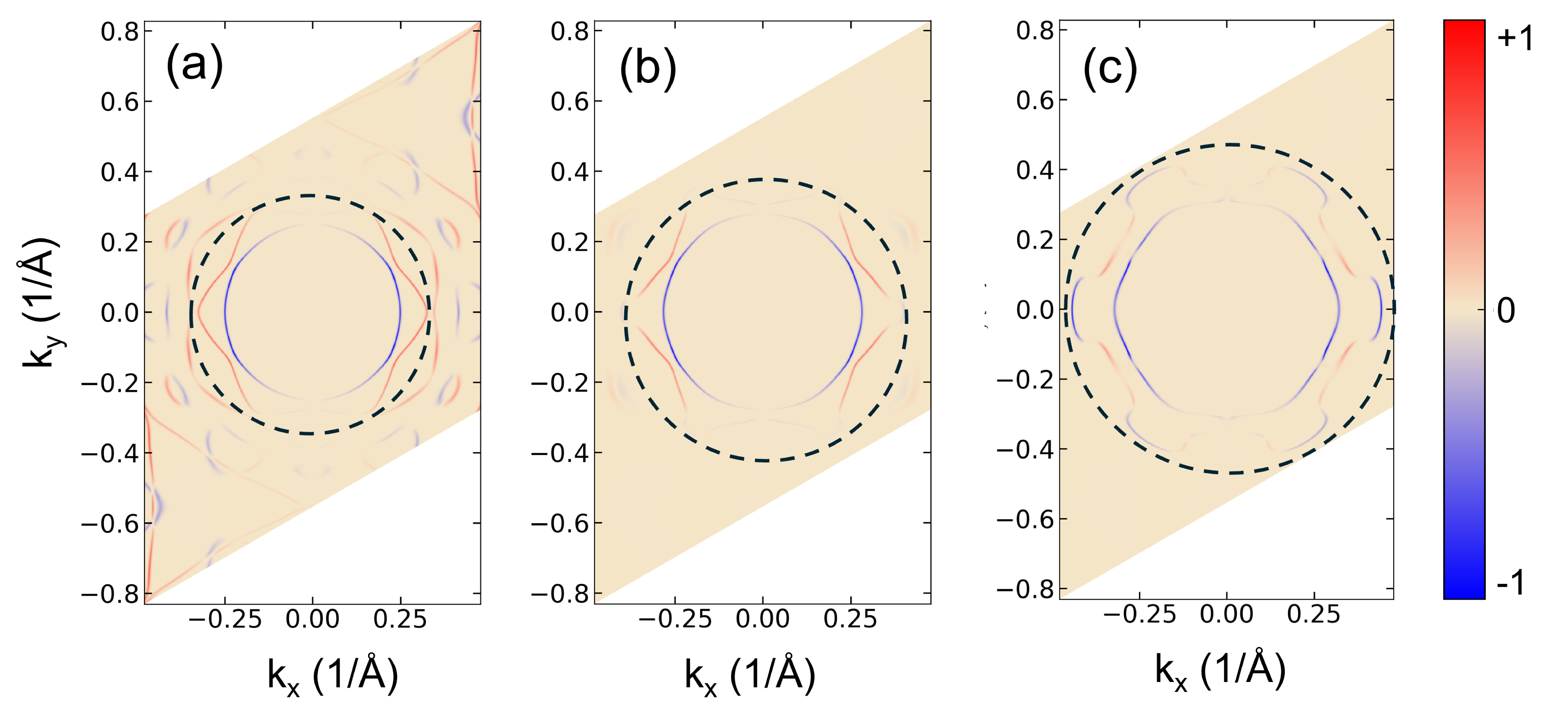}
\caption{\label{5} The $\mathbf{k}$-resolved EE coefficient ($\chi_{yx}(E_{F},\mathbf{k})$) for UU state of bilayer PtBi$_{2}$ at (a) $E_{F}$ = 0 eV, (b) $E_{F}$ = 0.1 eV, and (c) $E_{F}$ = 0.2 eV, respectively. The red and blue denotes the positive and negative values of $\chi_{yx}(E_{F},\mathbf{k})$, respectively, which have been normalized to the maxima value. The EE contributions states from inner electron pockets near the $\Gamma$ point are circled, while the states out of the circle are from outer hole pockets.}
\end{figure*}
The variation of the polarization state changes the crystal symmetry and electronic structure, which may further lead to distinct EE responses. Specifically, both interlayer-parallel UU and DD states belong to the $C_{3v}$ point group, which permits only two nonzero components of the EE tensor, satisfying $\chi_{yx}=-\chi_{xy}$. Switching the polarization from the UU to the DD state is equivalent to a spatial inversion operation and therefore reverses the sign of the EE coefficient. By contrast, the interlayer-antiparallel UD and DU states are centrosymmetric and belong to the $C_{2h}$ point group. Consequently, their EE response is forbidden by inversion symmetry. To gain a quantitative understanding of the EE, we calculate the symmetry-allowed EE coefficients as functions of chemical potential and polarization state, as shown in Fig. \ref{3}. It can be clearly seen that reversing the polarization between the two interlayer-parallel states, UU and DD, indeed switches the sign of the EE coefficient, whereas the interlayer-antiparallel UD and DU states exhibit zero EE, consistent with the above symmetry analysis. This demonstrates that the EE in bilayer PtBi$_{2}$ can be reversibly switched among positive, negative, and zero-response states through polarization-state control, enabling multistate manipulation of charge–spin conversion. For the UU state, the EE coefficient at $E_F$=0 eV is approximately $1.12\times{10}^{-9} \hbar \textup{m/V}$. Notably, such an EE can exist intrinsically in this metallic ferroelectric system without the need for additional doping. This behavior differs from that in previously reported ferroelectric insulators, where the EE can only be realized after doping the system into a metallic regime. When the Fermi level is shifted upward to $E_F$=0.05 eV, the EE coefficient reaches a local maximum of $1.54\times{10}^{-9} \hbar \textup{m/V}$. Further upward shifting of the Fermi level, which is in principle accessible by electrostatic gating, leads to a monotonic decrease and eventual sign reversal of the EE coefficient. At $E_F$=0.2 eV, the EE coefficient reaches $-1.54\times{10}^{-9} \hbar \textup{m/V}$. By normalizing the above EE coefficients with the calculated electrical conductivity and considering an effective thickness of 11 \AA, the corresponding normalized EE efficiencies are $6.54\times{10}^9$, $1.16\times{10}^{10}$, and $ -2.23\times{10}^{10} \hbar/(\mathrm{A}\cdot\mathrm{cm})$ at $E_F$=0 eV, 0.05 eV, and 0.2 eV, respectively. These magnitudes are comparable to, and even larger than, those of previously reported ferroelectric systems \cite{FE3,FE4,FE6,REE17, j5s5-m7j5}, which typically fall within the range of ${10}^9\sim{10}^{10} \hbar/(\mathrm{A}\cdot\mathrm{cm})$

Microscopically, the EE is dominated by electronic states at the Fermi surface. Therefore, examining the evolution of these electronic states with respect to the Fermi level is essential for elucidating the underlying mechanism of the nontrivial EE in the interlayer-parallel configurations. To this end, we analyze the Fermi surfaces states and the corresponding spin textures at different chemical potentials for the UU state of bilayer PtBi$_{2}$, as shown in Fig. \ref{4}. At $E_F$=0 eV, two electron pockets near the $\Gamma$ point exhibit Rashba-like spin textures with opposite helicities. The Fermi contour of the inner electron pocket is nearly circular, resembling that expected from typical Rashba-type spin splitting, whereas the outer electron pocket displays a warped Fermi contour, consistent with the $C_{3v}$ symmetry of the system. In addition, several outer hole pockets appear away from the $\Gamma$ point, which may also contribute to the EE. As the Fermi level is shifted to higher energies, the inner electron pockets expand in area while their shapes and spin textures remain qualitatively unchanged. Meanwhile, the outer hole pockets gradually shrink and eventually disappear at $E_F$=0.2 eV. The above analysis applies to the UU state. For the DD state, the Fermi-surface geometry remains unchanged over the entire energy range considered, whereas the spin textures are reversed due to the inversion-related polarization switching (not shown).

The evolution of the Fermi surface with the Fermi level leads to corresponding changes in its contribution to the EE. To gain further insight, we rewrite the EE into a sum of k-resolved expression: \begin{equation}
\begin{aligned}
    \chi_{yx} = \frac{(2\pi)^{3}}{VN} \sum_{k} \chi_{yx}(E_{F},\mathbf{k})
\end{aligned}
\end{equation}
where $V$ and $N$ is the volume of the unit cell and the total number of k points used to sample the Brillouin zone (BZ), respectively.
The distribution of the k-resolved EE in the UU configuration of bilayer PtBi$_{2}$ is shown in Fig. \ref{5}. At $E_F$=0 eV, the two inner electron pockets give rise to EE contributions with opposite signs, consistent with the opposite helicities of their Rashba-like spin textures. In addition, the outer hole pockets also make a sizable contribution to the EE, with the positive component being dominant. As the Fermi level is shifted upward from $E_F$=0 eV to 0.2 eV, these outer hole pockets gradually shrink and eventually disappear. Consequently, the positive contribution is strongly suppressed, and the total EE becomes increasingly dominated by the two inner electron pockets, whose combined contribution is negative. Meanwhile, the expansion of the inner electron pockets further enhances their negative EE contribution. At $E_F$=0.2 eV, an additional negative EE contribution emerges from the outer electron pocket, further strengthening the overall negative response. Together, these changes lead to a continuous decrease in the total EE and ultimately drive its sign reversal from positive to negative as the Fermi level is shifted upward. This analysis shows that the EE in the interlayer-parallel configuration of bilayer PtBi$_{2}$ is governed by the interplay between the inner electron pockets and the outer hole pockets. Such interplay can be tuned by shifting the Fermi level through electrostatic gating or chemical doping, thereby enabling sign reversal of the total EE.

\subsection{Spin Hall effect}

 \begin{figure*}[ht]
\includegraphics[scale = 0.33 ]{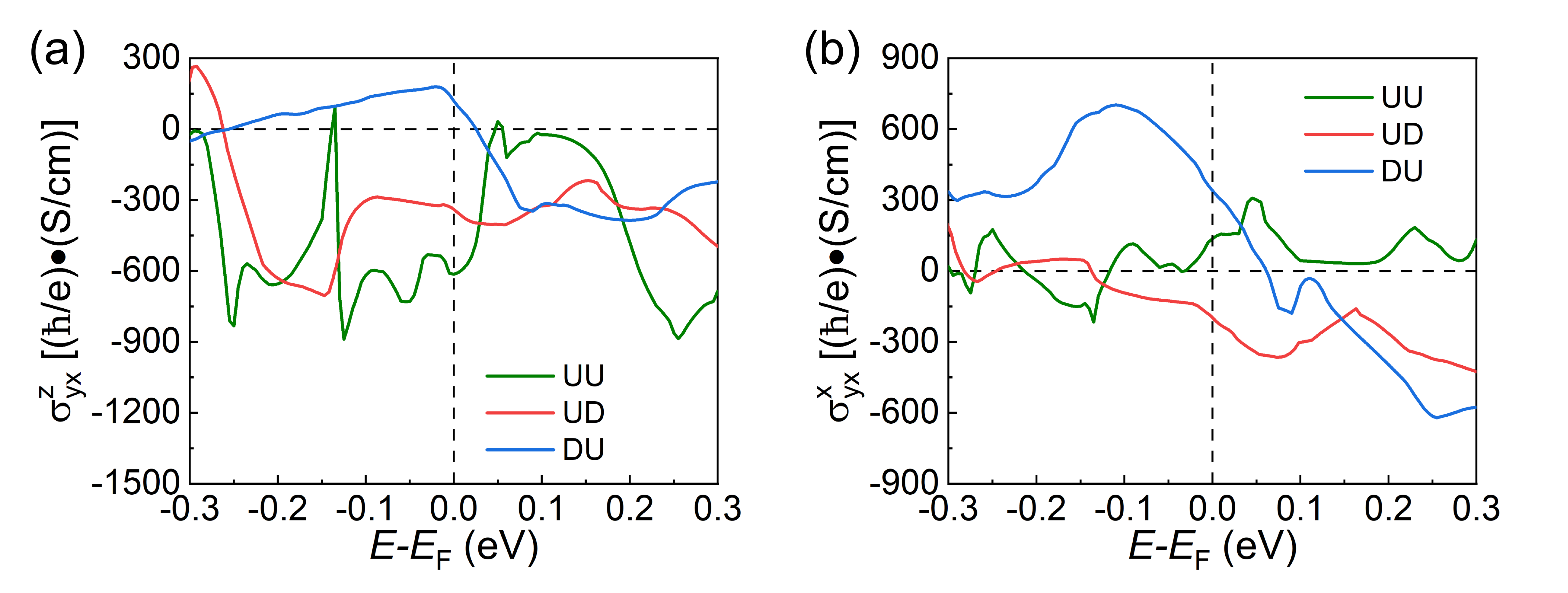}
\caption{\label{6} Calculated symmetry-allowed intrinsic spin Hall conductivity (a) $\sigma^{z}_{yx}$ and (b) $\sigma^{x}_{yx}$ in bilayer PtBi$_{2}$ with distinct polarization configurations as a function of Fermi energy. Note that DD configuration yield the identical numerical result with UU configuration due to the constraint of inversion symmetry operation, and is not shown here.}
\end{figure*}

Besides the nonvolatile manipulation of the Edelstein effect, another important charge-to-spin conversion phenomenon—the SHE—can also be controlled in bilayer PtBi$_{2}$ by switching the polarization state. Note that UU/DD states possess $C_{3v}$ symmetry, whereas the UD/DU states exhibit $C_{2h}$ symmetry. Under the constraint of these symmetries, both the conventional SHE component $\sigma^{z}_{yx}$ and the unconventional component $\sigma^{x}_{yx}$ are allowed, while $\sigma^{y}_{yx}$ is forbidden in either UU/DD state or UD/DU state by the mirror symmetry $M_y$. To obtain quantitative insight, we first evaluate $\sigma^{z}_{yx}$ in different polarization states of bilayer PtBi$_{2}$ as a function of the Fermi energy, as shown in Fig. \ref{6}(a). The results reveal that the SHE response varies substantially with the polarization configuration. Specifically, in the interlayer-parallel UU state, $\sigma^{z}_{yx}$ exhibits pronounced fluctuations as the Fermi level changes, alternating between sizable and nearly vanishing values. At $E_F$ = 0 eV, $\sigma^{z}_{yx}$ reaches a considerable value of of -610\ $(\hbar /e) \cdot $ (S/cm), which is comparable to, and even larger than, those of some previously reported ferroelectric systems \cite{PhysRevLett.133.146605,j5s5-m7j5,wang2020spin,slawinska2020ultrathin,fang2020tuning,xu2022regulating,chen2025quantum,GeTe2}, typically on the order of 100 \ $(\hbar /e) \cdot $ (S/cm). When the Fermi level is shifted upward to $E_F$ = 0.1 eV, the negative $\sigma^{z}_{yx}$ is strongly suppressed and decreases to a much smaller value of -25 \ $(\hbar /e) \cdot $ (S/cm). This suppression originates from the expansion of the positive contribution of the spin Berry curvature in momentum space, as can be seen in Fig. \ref{8} in Appendix D. For the UD state, $\sigma^{z}_{yx}$ is approximately -350 \ $(\hbar /e) \cdot $ (S/cm) at $E_F$ = 0 eV, and shifting the Fermi level upward to $E_F$ = 0.1 eV induces only a moderate change. In contrast, for the DU state, a positive $\sigma^{z}_{yx}$ of 118 \ $(\hbar /e) \cdot $ (S/cm) appears at $E_F$ = 0 eV. Upon increasing the Fermi level to $E_F$ = 0.1 eV, $\sigma^{z}_{yx}$ changes sign and becomes -318 \ $(\hbar /e) \cdot $ (S/cm). This sign reversal can be attributed to the reduction of the positive spin Berry curvature near the Brillouin-zone boundary, as shown in Fig. \ref{8} in Appendix D.

Having clarified the behavior of $\sigma^{z}_{yx}$, we next evaluate the unconventional SHE component $\sigma^{x}_{yx}$, as shown in Fig. \ref{6}(b). For the UU state, $\sigma^{x}_{yx}$ at $E_F$ = 0 eV is 136 \ $(\hbar /e) \cdot $ (S/cm), which is much smaller in the magnitude than the corresponding $\sigma^{z}_{yx}$ value of -610 \ $(\hbar /e) \cdot $ (S/cm). Shifting the Fermi level upward to $E_F$ = 0.1 eV leads only to a moderate variation in $\sigma^{x}_{yx}$. A similar trend is observed in the UD state, where $\sigma^{x}_{yx}$ at $E_F$ = 0 eV is -189 \ $(\hbar /e) \cdot $ (S/cm), smaller in magnitude than $\sigma^{z}_{yx}$ (-334 \ $(\hbar /e) \cdot $ (S/cm)), and remains only moderately affected by the upward shift of the Fermi level. For the DU state, however, $\sigma^{x}_{yx}$ reaches 340 \ $(\hbar /e) \cdot $ (S/cm) at $E_F$ = 0 eV, exceeding the corresponding $\sigma^{z}_{yx}$ value of 118 \ $(\hbar /e) \cdot $ (S/cm). When the Fermi level is shifted upward to $E_F$ = 0.1 eV, $\sigma^{x}_{yx}$ decreases and changes sign to -64 \ $(\hbar /e) \cdot $ (S/cm). Microscopically, this sign change can be understood as resulting from the shrinkage of the positive spin Berry curvature in the Brillouin zone, as can be seen in the Fig. \ref{9} in Appendix D. These results demonstrate that switching the polarization state in bilayer PtBi$_{2}$ can induce distinct SHE magnitudes and Fermi-level dependence. This provides a promising route for the nonvolatile control of spin currents in ferroelectric systems.

\section{Conclusions}
In conclusion, we propose a mechanism for the non-volatile control of both the EE and the SHE in ferroelectric systems. By stacking two ferroelectric monolayers into a bilayer, both interlayer-parallel and interlayer-antiparallel polarization configurations can be realized. This provides an additional degree of freedom for manipulating charge–spin conversion. In the interlayer-parallel configurations, the sign of the EE can be reversed by switching the total polarization. In contrast, by changing the polarization configuration from interlayer-parallel to interlayer-antiparallel, the EE can be switched off, enabling multistate control of current-induced spin accumulation. Moreover, the magnitude of the SHE can also be non-volatilely tuned by changing the polarization configuration of the ferroelectric bilayer between interlayer-parallel and interlayer-antiparallel. 

Based on first-principles calculations, we demonstrate that such multistate-controllable intrinsic EE and SHE can be realized in bilayer ferroelectric PtBi$_{2}$, where both interlayer-parallel and interlayer-antiparallel configurations are energetically stable. Microscopically, the EE in the interlayer-parallel states of bilayer PtBi$_{2}$ arises from the competition between the negative contribution from the inner electron pockets and the positive contribution from the outer hole pockets. This competition can be modified by shifting the Fermi level upward, leading to a sign reversal of the total EE. Furthermore, the SHE originates from the spin Berry curvature, whose distribution in k-space can be altered either by switching between interlayer-parallel and interlayer-antiparallel polarization configurations or by shifting the Fermi level, thereby resulting in a tunable SHE. 

Our study thus not only provides new insights 
on the functional potential of multistate ferroelectric systems, but also paves the way towards non-volatile and multistate control of charge–spin conversion in realistic materials, which is beneficial for electrically controllable spintronic devices in the future.

\section{Acknowledgments}
This project was supported by the European Union Graphene Flagship project 2DSPIN-TECH (grant agreement No. 101135853) and SFB 1277 (Project-ID 314695032). W.P. and X.J. contributed equally to this work. 

\section{Appendix A: Computational methods}
Our structural relaxation is carried out using density functional theory (DFT) which is implemented in VASP\cite{VASP} package based on the projector augmented waves (PAW)\cite{PAW} in the generalized-gradient approximation (GGA). The Perdew-Burke-Ernzerhof (PBE) functional is adopted to describe the electron exchange-correlation effect\cite{PBE}. The structures of bilayer PtBi$_2$ are relaxed until the forces are less than 0.001 eV/\AA and the energy difference is less than $10^{-6}$ eV. The energy cutoff is set to be 400 eV, and the k-space is sampled with a Gamma-centered $12\times12\times1$ k-mesh. The DFT-D3 van der Waals correction is adopted during the relaxation\cite{DFT-D3}. A vacuum layer of larger than 20 \AA is adopted to decouple the periodic layers along $z$ direction.
The calculated in-plane lattice constant for four distinct polarization states (UU/DD, UD, DU) are all being approximately 6.56 \AA, which is consistent with the reported value in monolayer PtBi$_{2}$\cite{j5s5-m7j5}. Meanwhile, the calculated interlayer distance between between upper and lower Bi atoms at the interface for UU/DD, UD and DU states are 2.27 \AA, 1.86 \AA and 2.87 \AA, respectively. The ferroelectric switching path ways between distinct polarization configurations are calculated with the nudged elastic band (NEB) approach\cite{NEB}. Having obtained the optimized atomic structure of bilayer PtBi$_{2}$, we calculate the electronic structure using Quantum Espresso code\cite{QE} with fully relativistic pseudopotentials with PAW method and PBE functional. The energy cutoff of 500 Ry for charge density and 60 Ry for wave function is adopted during the self-consistent calculation.

After the DFT calculations, the Wannier Hamiltonian is constructed using Wannier90 code\cite{w901} in conjunction with Quantum Espresso code. During the wannierization procedure, we select the Pt-$d$ and Bi-$p$ orbitals as the projection basis. Based on the obtained wannier Hamiltonian, we adopt the Wanniertools package to evaluate the spectral density of Fermi surface as well as the corresponding spin textures\cite{wtools}. For the calculation of EE coefficient $\chi_{ij}$, we employ the Linres code\cite{linears} based on the Kubo formula. Specifically, the formula of $\chi_{ij}$ can be written as: 
\cite{Kubo1,Kubo2,PhysRevB.110.214419}:
\begin{equation}
    \chi_{ij} = -\frac{e\hbar}{\pi V N} \sum_{\mathbf{k},m,n} \frac{\Gamma^{2} \textup{Re}(\langle n\mathbf{k} |\hat{S}_{i} | m\mathbf{k} \rangle \langle m\mathbf{k} | \hat{v}_{j} | n\mathbf{k} \rangle  )  }{[(E_{f}-\epsilon_{n\mathbf{k}})^{2}+ \Gamma^{2}][(E_{f}-\epsilon_{m\mathbf{k}})^{2}+ \Gamma^{2}]}
\end{equation}
 
Here, $e$ is the elementary charge; $n$ and $m$ denote band indices; $\textbf{k}$ is the Bloch vector; $E_{F}$ is the Fermi energy; $\hat{v}$ is the velocity operator, $\hat{S}$ is the spin operator; $\epsilon_{n\textbf{k}}$ is the eigenvalue; $V$ is the volume of unit cell; $N$ is the total number of $k$ points used to sample the BZ; and $\Gamma$ is the disorder parameter, which is related to the relaxation time $\tau$ through $\tau = \hbar/2\Gamma$. 
It should be noted that such EE originates mainly from electronic states at the Fermi surface. This can be understood more clearly in the limit of a small broadening parameter $\Gamma$, where the corresponding expression can be approximated as\cite{Kubo1}:

\begin{equation}
    \chi_{ij} = -\frac{e\hbar}{2 \Gamma  N} \sum_{\textbf{k},n}     \delta(\epsilon_{nk}- E_{F}) 
 \langle n\textbf{k} | \hat{S}_{i} | n\textbf{k} \rangle \langle n\textbf{k} |\hat{v}_{j} | n\textbf{k} \rangle  
\end{equation}
where the $\delta$-function explicitly indicates that only states at the Fermi level contribute to the EE.
Here we set $\Gamma$= 0.01 eV, which roughly corresponds to a reasonable sub-100 fs scattering rate (broadening) while remaining much smaller than interband separations. A $600 \times 600 \times 1$ k-point mesh is used to obtain the converged EE coefficient.

The SHE consists of both intrinsic and extrinsic contributions. The intrinsic contribution is commonly described by the Kubo formula, whereas the extrinsic contribution arises from skew-scattering and side-jump mechanisms and is therefore strongly dependent on disorder\cite{SHE}. In this work, we focus on the intrinsic SHE in a weak scattering limit. The intrinsic spin Hall conductivity $\sigma^{i}_{yx}$ ($i = x,y,z$) is calculated using the Kubo formula implemented in the Wannier90 code\cite{PhysRevB.98.214402,PhysRevLett.94.226601}: 
\[
\begin{aligned}
\sigma^{i}_{yx}
&=
-\frac{e\hbar}{V N}
\sum_{\mathbf{k},n}
f_{n\mathbf{k}}
\sum_{m\neq n}
\frac{
2\,\mathrm{Im}
\left[
\left\langle n\mathbf{k} \middle| \hat{j}^{i}_{y} \middle| m\mathbf{k} \right\rangle
\left\langle m\mathbf{k} \middle| \hat{v}_{x} \middle| n\mathbf{k} \right\rangle
\right]
}{
\left(\epsilon_{n\mathbf{k}}-\epsilon_{m\mathbf{k}}\right)^2
},
\end{aligned}
\]
where $\hat{j}^{i}_{y}
=
\frac{1}{2}
\left\{
\hat{s}_{i}, \hat{v}_{y}
\right\}$ is the spin current operator. 
Such SHE is intrinsic and does not depend on the extrinsic disorder parameter $\Gamma$. 
Here, the SHE is calculated using a $600 \times 600 \times 1$ k-point mesh to arrive at the converged results. 

\section{Appendix B: Switching the polarization states}
 \begin{figure}[ht]
\includegraphics[scale = 0.2 ]{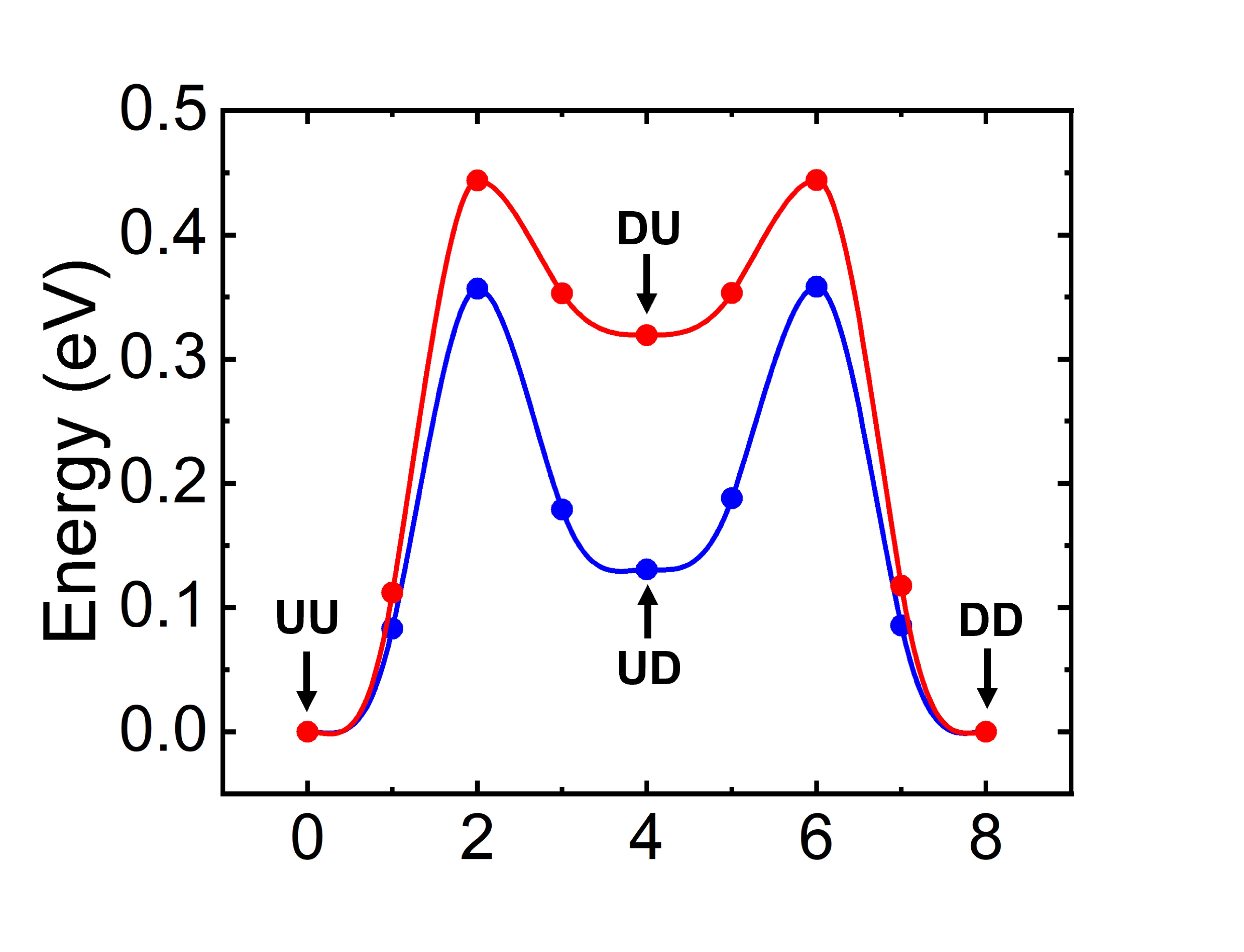}
\caption{\label{7} Transition pathway between distinct polarization configurations in bilayer PtBi$_2$. The energy per unit cell is shown on the $y$-axis. The $x$-axis represents the reaction coordinate along the polarization switching pathway, and each dot corresponds to an intermediate atomic configuration. Four metastable states, namely UU, UD, DU, and DD, whose atomic structure can be seen in Fig. \ref{2}, are labeled in the figure.}
\end{figure}

We show the kinetic switching pathways between distinct polarization states in Fig. \ref{7}. By using the method adopted previously in monolayer PtBi$_2$\cite{Wu}, the magnitude of out-of-plane electric polarization for bilayer PtBi$_{2}$ in UU/DD phase is calculated to be 0.86 pC/m, which is comparable with various typical 2D ferroelectric systems, such as CuCrSe$_{2}$ (1.9 pC/m)\cite{10.1093/nsr/nwz169}, bilayer MoS$_{2}$ (0.61 pC/m)\cite{PhysRevLett.112.157601}, and 1T'-WTe$_{2}$ (0.35 pC/m)\cite{yang2018origin}. Formally, one can estimate the critical switching field by $E_{c} = \Delta E_{\textup{barrier}}/P$, in which $\Delta E_{\textup{barrier}}$ denotes the switching barrier and $P$ denotes the out-of-plane electric dipole in a unit cell. In our case, the critical field for switching from interlayer-parallel to interlayer-antiparallel states is thus estimated to be at the order of 100 V/nm, hardly realizable experimentally. However, such an estimated switching field should be regarded as an intrinsic upper bound for a coherent, unit-cell-level polarization reversal. In experiment, switching is more likely mediated by nucleation of reversed domains followed by domain-wall propagation, so the applied field only needs to overcome a local nucleation barrier rather than the full homogeneous switching barrier, thus lead to a much smaller critical field\cite{PhysRevB.108.024305,shin2007nucleation}. A similar situation is revealed in monolayer ferroelectric metal CuCrSe$_{2}$. Based on its calculated switching barrier and electric polarization\cite{10.1093/nsr/nwz169}, the estimated intrinsic critical field is also on the order of 100 V/nm, yet experimental piezoresponse force microscopy measurements demonstrate polarization switching at a much smaller bias/field scale (less than 5V/nm)\cite{sun2024evidence}. In fact, the polarization reversal of 2D ferroelectric metals has also been experimentally reported previously in bilayer 1T'-WTe$_{2}$ under a very small electric field (less than 1V/nm)\cite{FE9}. Given that bilayer PtBi$_{2}$ has a comparable switching barrier and out-of-plane polarization to monolayer CuCrSe$_{2}$, it is plausible that the polarization state of bilayer PtBi$_{2}$ can also be switched experimentally under a realistic electric field in the range of 1-10 V/nm.

\section{Appendix C: Normalizing the EE coefficient}
\begin{figure*}[ht]
\includegraphics[scale = 0.46 ]{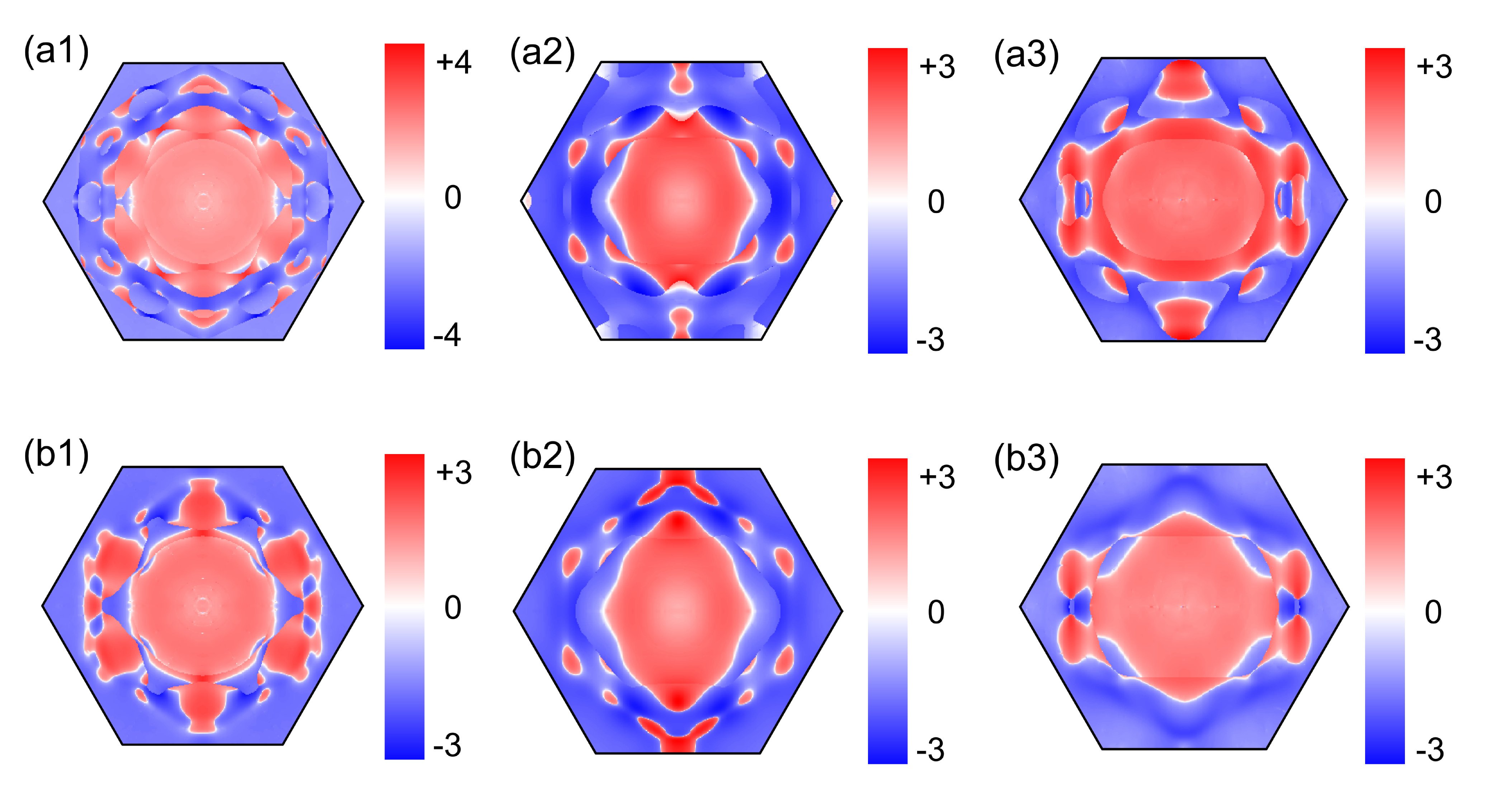}
\caption{\label{8} Spin Berry curvature distribution  for $\Omega^{z}_{yx}$ of bilayer PtBi$_{2}$ in first BZ. (a1)-(a3) shows the spin Berry curvature in UU, UD and DU state at $E_F$ = 0 eV, respectively. (b1)-(b3) shows the spin Berry curvature in UU, UD and DU state at $E_F$ = 0.1 eV, respectively.  The unit of spin Berry curvature is \AA$^2$.  }
\end{figure*}

\begin{figure*}[ht]
\includegraphics[scale = 0.46 ]{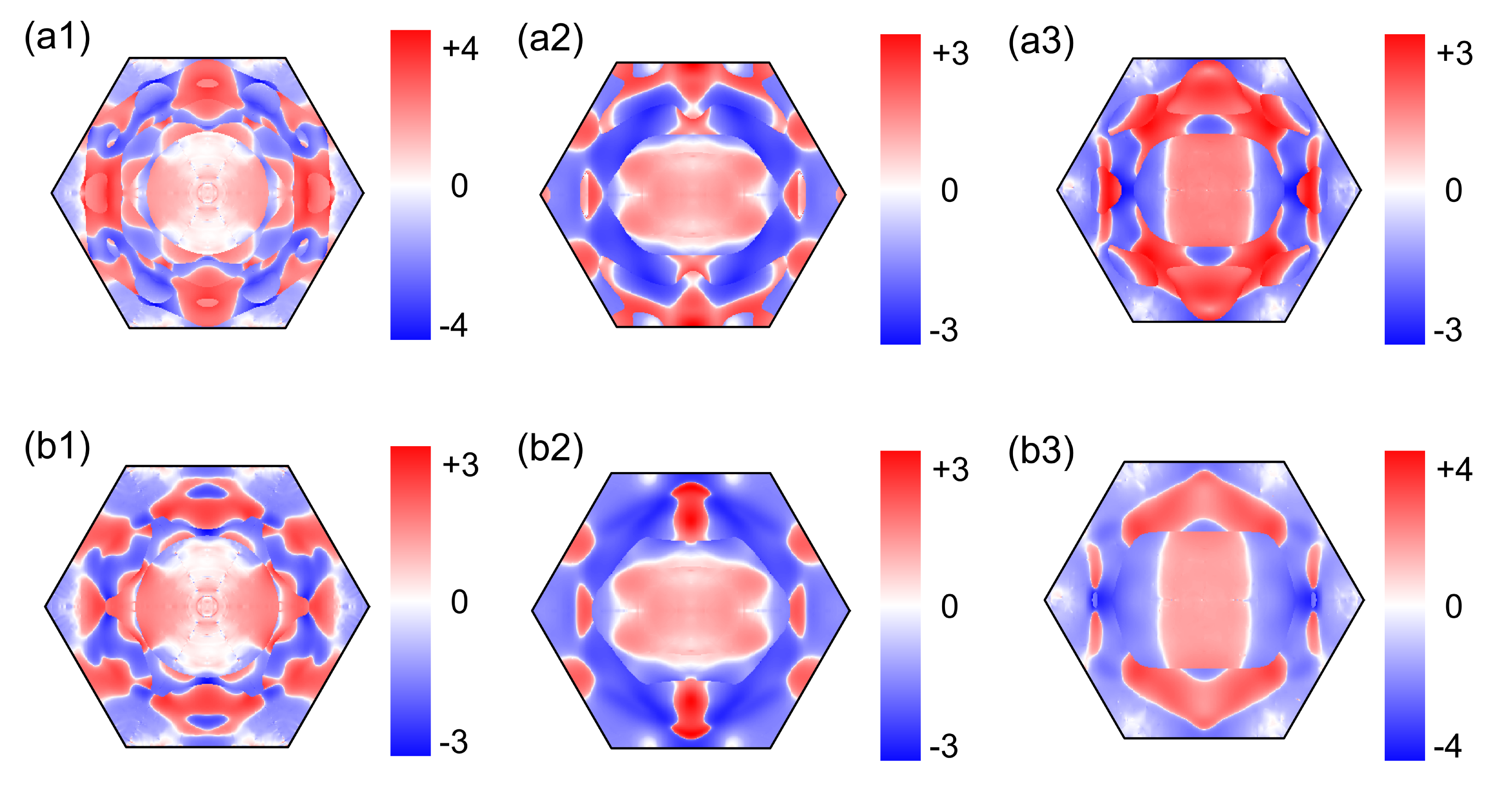}
\caption{\label{9} Spin Berry curvature distribution  for $\Omega^{x}_{yx}$ of bilayer PtBi$_{2}$ in first BZ. (a1)-(a3) shows the spin Berry curvature in UU, UD and DU state at $E_F$ = 0 eV, respectively. (b1)-(b3) shows the spin Berry curvature in UU, UD and DU state at $E_F$ = 0.1 eV, respectively.  The unit of spin Berry curvature is \AA$^2$.  }
\end{figure*}

In addition to the Edelstein effect discussed in the main text,
\begin{equation}
    \delta S_i = \chi_{ij} E_j ,
\end{equation}
where $\chi_{ij}$ characterizes the spin angular momentum, in units of $\hbar$, induced in a single unit cell by an electric field of $1~\mathrm{V/\AA}$ and has the unit of $\hbar~\mathrm{\AA/V}$, one may also define a normalized current-induced Edelstein effect as
\begin{equation}
    \delta S_i = \chi'_{ij} J_j .
\end{equation}
Here, $\chi'_{ij}$ denotes the three-dimensional volume density of spin angular momentum, in units of $\hbar/\mathrm{cm}^{3}$, induced by a current density of $1~\mathrm{A/cm}^{2}$. Therefore, $\chi'_{ij}$ has the unit of $\hbar/(\mathrm{A}\cdot\mathrm{cm})$.

The electric current density is related to the electric field through
\begin{equation}
    J_j = \sigma_{jj} E_j ,
\end{equation}
where $\sigma_{jj}$ is the longitudinal conductivity along the $j$ direction. Accordingly, the Edelstein response coefficient $\chi_{ij}$ can be converted into the normalized current-induced Edelstein coefficient $\chi'_{ij}$ as
\begin{equation}
    \chi'_{ij} = \frac{\chi_{ij}}{V \sigma_{jj}} ,
\end{equation}
where $V$ is the effective volume of the system.

The longitudinal conductivity $\sigma_{jj}$ is calculated using the Kubo formula implemented in the \textsc{Linres}\cite{linears} code:
\begin{equation}
    \sigma_{jj}
    =
    \frac{e^{2}\hbar}{\pi V N}
    \sum_{\mathbf{k},m,n}
    \frac{
    \Gamma^{2}
    \operatorname{Re}
    \left[
    \langle n\mathbf{k} |\hat{v}_{j}| m\mathbf{k} \rangle
    \langle m\mathbf{k} |\hat{v}_{j}| n\mathbf{k} \rangle
    \right]
    }
    {
    \left[(E_F-\epsilon_{n\mathbf{k}})^2+\Gamma^2\right]
    \left[(E_F-\epsilon_{m\mathbf{k}})^2+\Gamma^2\right]
    } .
\end{equation}

Here, $e$ is the elementary charge; $n$ and $m$ denote band indices; $\textbf{k}$ is the Bloch vector; $E_{F}$ is the Fermi energy; $\hat{v}$ is the velocity operator, $\hat{S}$ is the spin operator; $\epsilon_{n\textbf{k}}$ is the eigenvalue; $V$ is the volume of unit cell; $N$ is the total number of k points used to sample the BZ; and $\Gamma$ = 0.01 eV is the disorder parameter. Using the calculated longitudinal conductivity and an effective bilayer thickness of approximately 11 \AA, we obtain the normalized Edelstein efficiency. This normalized quantity is largely insensitive to the choice of broadening parameter and therefore provides a reliable reference for comparing the current-induced Edelstein response.

\section{Appendix D: k-resolved Spin Hall effect}
 
To gain a better understanding on the microscopic origin of SHE, we rewrite the spin Hall conductivity as:
    \begin{equation}
    \sigma_{yx}^{i}
    =
    \frac{e}{2 V N}
    \sum_{\mathbf{k}}
    \Omega_{yx}^{i}(\mathbf{k}) .
\end{equation}
Here $\Omega_{yx}^{i}(\mathbf{k})$ ($i = x,z$) is usually termed as spin Berry curvature, which can be further expressed as:
\begin{equation}
\begin{aligned}
  &  \Omega_{yx}^{i}(\mathbf{k})
    =
    \sum_{n}
    f_{n\mathbf{k}}
    \Omega_{n,yx}^{i}(\mathbf{k}) \\
 &   =
    \sum_{n}
    f_{n\mathbf{k}}
    \hbar^{2}
    \sum_{m\neq n}
    \frac{
    -2\,\mathrm{Im}
    \left[
    \left\langle
    n\mathbf{k}
    \left|
    \frac{1}{2}
    \left\{
    \hat{\sigma}_{i},
    \hat{v}_{y}
    \right\}
    \right|
    m\mathbf{k}
    \right\rangle
    \left\langle m\mathbf{k}
    \left|
    \hat{v}_{x}
    \right|
    n\mathbf{k}
    \right\rangle
    \right]
    }
    {
    \left(
    \epsilon_{n\mathbf{k}}
    -
    \epsilon_{m\mathbf{k}}
    \right)^{2} 
    } .
\end{aligned}
\end{equation}

We take the logarithm of $\Omega_{yx}^{i}(\mathbf{k})$ when plotting it, which allows the rapid variations in the spin Berry curvature to be visualized more clearly. The logarithmic quantity is defined as:
\begin{equation}
    \Omega'
    =
    \begin{cases}
        \operatorname{sgn}(\Omega)\log_{10}|\Omega|, & |\Omega| > 10, \\[4pt]
        \Omega/10, & |\Omega| \leq 10 .
    \end{cases}
\end{equation}

The results for $\Omega^{z}_{xy}$ are shown in Fig. \ref{8}. As can be qualitatively observed, at $E_F$=0 eV, the DU state exhibits a more positive spin Berry curvature distribution near the boundary of the BZ. This behavior differs from those of the UU and UD states, where the positive spin Berry curvature is mainly concentrated around the $\Gamma$ point. This observation is consistent with the fact that, at $E_F$=0 eV, the UU and UD states possess more negative SHE values than the DU state. When the Fermi level is shifted to $E_F$=0.1 eV, a more positive spin Berry curvature distribution emerges near the BZ boundary for the UU state, whereas the distributions for the UD states show no significant qualitative changes. Meanwhile, more negative spin Berry curvature emerges near the BZ boundary for DU states. The above analysis is consistent with the fact that shifting $E_F$ from 0 to 0.1 eV leads to an increase in the $\sigma^{z}_{yx}$ of the UU state and decrease in the $\sigma^{z}_{yx}$ of the DU state , while inducing only moderate changes in the UD state. 

We also show the results for $\Omega^{x}_{xy}$ in Fig. \ref{9}.  As can be seen, at $E_{F}$ = 0 eV, the DU state exhibits more positive spin Berry curvature contribution than that of UU and UD states. This observation is consistent with the fact that, at $E_F$=0 eV, the DU state possess more positive $\sigma^{x}_{yx}$ than that of UU and UD states. When the Fermi level is shifted to $E_{F}$ = 0.1 eV, the positive spin Berry curvature contributions are suppressed in DU state, which is consistent with the fact that shifting the Fermi level from 0 to 0.1 eV leads to an decrease in the $\sigma^{x}_{yx}$ of the DU state. 

%


\end{document}